\documentclass[12pt]{iopart}
\usepackage{iopams}  
\usepackage{graphicx}

\begin{document}

\title[]{Stimulation of static deconfined medium by multiple hard partons}

\author{Martin Schulc$^1$ and Boris Tom\'a\v{s}ik$^{1,2}$}

\address{$^1$Czech Technical University in Prague, FNSPE, 11519 Prague, Czech Republic}
\address{$^2$Univerzita Mateja Bela, 97401 Bansk\'a Bystrica, Slovakia}
\eads{\mailto{boris.tomasik@umb.sk}}

\begin{abstract}
We investigate the response of non-expanding deconfined hot matter to energy and 
momentum deposition from a pair of partons moving with high energies. Several 
situations are examined with partons moving so that the generated wakes in the medium
interact. The resulting energy and flow profiles are studied. Such cases are relevant for 
nuclear collisions at the LHC where several hard partons are produced in a single collision
and their contribution to collective expansion of the fireball may be important.  
\end{abstract}



\section{Introduction}
\label{s:intro}

Nuclear collisions prepared at the Large Hadron Collider posses several unique features. 
Apart from the highest energy density ever obtained in a lab the initial state of the hot 
and dense strongly interacting medium includes a number of hard partons. 
For example, in a central Pb+Pb collision at full LHC energy  of 5.5~$A$ TeV 
one expects on average more than 8 hard partons with transverse energy above 20 GeV
within central 5 units of pseudorapidity \cite{Accardi:2004gp}. 
This number grows exponentially if we decrease the required energy of the jets.

Such hard partons deposit a large part (if not all) 
of their energy and momentum into the thermalised medium. This 
leads to lumps of energy and momentum density in the hot matter. Currently, much 
attention is devoted to such lumps in the energy density---called sometimes hot 
spots---which appear in the initial state for the hydrodynamical 
expansion 
\cite{Gyulassy:1996br,Socolowski:2004hw,Andrade:2008xh,Petersen:2009vx,%
Petersen:2010zt,Holopainen:2010gz,Schenke:2011bn,Qiu:2011iv, Gardim:2012yp,%
Bozek:2012fw}. 
These initial 
state inhomogeneities are being linked with the measured azimuthal anisotropies of  
hadron distributions \cite{Florchinger:2011qf}. 
In this paper, however, we focus our attention to energy and momentum deposition 
from hard partons \cite{Betz:2008ka,Betz:2010qh,Betz:2010tb}
which goes beyond the hot spots picture in at least two aspects: 
Firstly, momentum transferred to the 
medium has a direction in contrast to the energy density making up a hot spot. Secondly, 
it is being deposited over some period of time and not just instantaneously at the beginning
of the expansion. Technically, this means that energy and momentum deposition 
from hard partons are not treated as an initial condition for the hydrodynamic 
expansion but rather through a source term of the hydrodynamic equations. 

Azimuths of the hard partons are distributed symmetrically. Nevertheless, their numbers 
in individual collisions are not  large and thus on event-by-event basis one might 
naturally expect anisotropies in the transverse flow. 
They are naturally expressed through higher order terms 
in the harmonic expansion in the azimuthal particle distribution. 
When  summed over a large number of events one would 
first expect any anisotropies in particle production to even out due to original 
isotropy of hard parton production. It has been, however, argued  
that due to collective response of the hydrodynamic medium an asymmetry of 
the transverse expansion  might be generated in the fireball that leads to quadrupole 
anisotropy in hadron production \cite{Tomasik:2011xn}. 
In that paper, however, the argument was supported 
only by a toy model simulation without really modelling  the hydrodynamic response
to the hard partons. 

It has been studied by several authors by now how the medium responds to 
energy and momentum deposition from one hard parton 
\cite{Betz:2008ka,Betz:2010qh,Betz:2010tb,Satarov:2005mv,CasalderreySolana:2004qm,%
Koch:2005sx,Ruppert:2005uz,%
Renk:2005si,Neufeld:2008hs,Neufeld:2010tz,Shuryak:2011vf,Bouras:2012mh}.
It has been also shown \cite{Betz:2008ka} that the generated wake continues 
to travel even after the hard parton lost all of its energy and thermalised in 
the medium. In this paper we investigate how the excitations of 
the hot matter continue to evolve in case there are two of them  generated. 
We confirm the ansatz made in the toy model \cite{Tomasik:2011xn} that two 
wakes merge when they approach each other and continue to stream united according 
to energy and momentum conservation. This validates the conclusions made in 
that paper that the effect contributes to the elliptic flow.

The rest of the paper is divided into three sections: First we introduce the hydrodynamic 
model that was used in our simulations. In Section \ref{s:res} we present results obtained 
with this model for simulations of energy and momentum deposition from pairs of jets 
in different configurations. Our findings are summarised in Section \ref{s:summ}.


\section{The hydrodynamic model}
\label{s:mod}

The relativistic hydrodynamic equations express the conservation  
of energy, momentum, and baryon number:
\begin{eqnarray}
\partial_{\mu} T^{\mu \nu}(x)=J^\nu\,  ,  \label{eq:energy} \\ 
\partial_{\mu} J^{\mu}_{B}(x)=0 \,  , \label{eq:baryon}
\end{eqnarray}
where $T^{\mu \nu}(x)$ is the energy-momentum tensor and $J^{\mu}_{B}(x)$ is the baryon current. The source current $J^\nu$ will be discussed below. In the 
case of relativistic ideal fluid, the energy-momentum tensor and baryon current are given by
\begin{eqnarray}
T^{\mu\nu} = (\epsilon + p)u^{\mu}u^{\nu}-pg^{\mu\nu} ,\\
J^{\mu}_{B}(x) = n_{B}(x)u^{\mu}(x),
\end{eqnarray}
where $\epsilon(x)$, $p(x)$, $n_{B}(x)$, and $u^{\mu}=\gamma(1,\vec{v}) $ are the 
proper energy density, pressure and baryon density which are evaluated in the rest frame of the 
fluid, and flow four-velocity, respectively.
The factor $ \gamma=(1-\vec{v}^{2})^{-\frac{1}{2}}$. 
We set the net baryon density to zero in this study and consider  medium without collective 
velocity field generated by pressure. 
The system of equations (\ref{eq:energy}) is closed by specifying the equation of state 
$p = p(\epsilon)$. 
In our study, we will employ the equation of state by Laine and Schroder \cite{Laine}, which
is derived from  high-order weak-coupling perturbative QCD calculation 
at high temperatures, a hadron resonance gas at low temperatures, and an analytic 
interpolation in the crossover region between the high and low temperatures. 

The term $J^\nu$ on the right-hand side of eq.~(\ref{eq:energy}) is the source term 
of energy-momentum conservation. It parametrises the deposition 
energy and momentum into the medium%
\footnote{With the same term one can describe 
extraction of energy and momentum from the medium, but this case is not 
relevant and thus not elaborated here.}.
In our study it will represent deposition from hard partons which are precursors of the 
jets. We will loosely call them jets although they have not yet built up their showers. 
The form of source term which describes the interaction of the jet with QGP is 
not known yet exactly, though some groups made progress on this topic \cite{Neufeld:2008hs,Bouras:2012mh,Friess,Gubser,Chesler,Renk,Renketal}. 
In covariant notation it is given by \cite{Betz:2008ka}
\begin{equation}
 J^{\nu}(x) = \sum_i \int_{\tau_{i,i}}^{\tau_{f,i}}d \tau \frac{dP^{\nu}_i}{d\tau}\delta^{4}(x^{\mu}-x^{\mu}_{jet,i}),
\end{equation}
where $\tau_{f,i}-\tau_{i,i}$ denotes the proper time interval associated with the evolution of 
the $i$-th hard parton, $\vec{x}_{jet,i}$ describes its position, and $dP^{\nu}_i/d\tau$ is its energy-momentum loss rate along its 
trajectory  $x^{\mu}_{jet,i}(\tau)=x^{\mu}_{0,i}+u^{\mu}_{jet,i}\tau$. Summation runs over 
all hard partons in the system. 
In practical simulation it is assumed that the jet deposits its energy and momentum over 
some region characterised by a Gaussian profile. Then, 
in non-covariant notation, the  source term that we use is 
\begin{equation}
J^\nu(x) = 
\sum_i \frac{1}{(\sqrt{2\pi}\sigma)^{3}}\exp \left ( {-\frac{[\vec{x}-\vec x_{jet,i}(t)]^{2}}{2\sigma^{2}}} \right ) \left ( \frac{dE_i}{dt},\frac{d\vec P_i}{dt} \right ), 
\end{equation}
where  it is chosen that $\sigma=0.3$ fm. 
Energy deposition and momentum deposition in the direction of the moving jet
are denoted $dE/dt$ and $dP/dt$,  respectively.
The system with jets evolves in time until the simulation is stopped. 
The energy loss is modelled according to simplified Bethe-Bloch prescription \cite{Bethe:1930} with an explosive burst 
of energy and momentum known as the Bragg peak \cite{Bragg}. Due to its interaction 
with the plasma, the jet will decelerate and its energy and momentum loss will change. 
The hard partons deposit energy and also momentum in the direction of their motion.
The momemtum loss is given as
\begin{equation}
\frac{dP}{dx} = \frac{1}{v_{jet}}\frac{dE}{dx} = a\frac{1}{v_{jet}^{2}},\label{eq:deposition}
\end{equation}
where $v_{jet}$ is the jet velocity and $a$ determines the absolute scale 
of the jet stopping. This equation shows that when the jet decelerates, the energy-momentum 
deposition increases and has a peak for $v_{jet}\rightarrow 0$. 
In order to determine the actual velocity of the jet one can introduce jet rapidity
\begin{equation}
y_{jet} = \frac{1}{2} \ln \frac{1+v_{jet}}{1-v_{jet}}\, ,
\end{equation}
and then 
use the ansatz (\ref{eq:deposition}) 
and the identity $dP/dy_{jet}=m \cosh y_{jet}$, to derive  the 
dependence of the time on the jet rapidity \cite{Betz:2008ka}
\begin{equation}
 t(y_{jet})=\frac{m}{a}\left [ \sinh y_{jet}-\sinh y_{0}-\arccos \frac{1}{\cosh y_{jet}}+\arccos \frac{1}{\cosh y_{0}} \right ]\,  ,
\end{equation}
where $y_{0}$ is the initial rapidity of the jet. 
For all simulations the initial velocity  was set to $v_{0}=0.9999$, the mass of the moving parton is assumed to be of the order of the constituent quark 
mass and was set to $m=0.3$ GeV. The initial energy loss rate was usually set to 
$a = -4.148$ GeV/fm (unless stated otherwise). This value was determined from 
the fact that in our simulation 
jet stops after $\Delta \tau = 5.0$ fm/$c$. The unperturbed static energy density of 
the medium was adjusted to $ \epsilon_{0}=20.0$ GeV/fm$^{3}$. 
The main difference between the ansatz described here and the Bethe-Bloch equation is that the momentum deposition is longitudinal rather than transverse.

For hydrodynamic simulation we use SHASTA algorithm \cite{Boris} together with 
a multidimensional flux-correction \cite{DeVore} which is an improved version of 
Zalesak's method
\cite{Zalesak}, to solve the (3+1)-dimensional system of hydrodynamic equations with source terms, e.g.\ eq.~(\ref{eq:energy}).


\section{Results}
\label{s:res}

We  examine various types of 
static medium responses to perturbation by moving jets. 
In order to compare our results to \cite{Betz:2008ka} we begin with excitation of QGP by 
one hard parton moving along the $x$-axis. In \cite{Betz:2008ka} this situation was 
simulated for lower energy densities and energy deposition. Our values are more
relevant for quark matter produced at the LHC while that paper was focussed more at RHIC%
\footnote{In \cite{Betz:2008ka} the authors
used  energy-momentum deposition $dE/dx=1.5$~GeV/fm and 
static quark gluon plasma with temperature $T_{0}=200$ MeV.}. 
The difference is also in the implemented equation of state: we use the  QCD 
parametrisation of \cite{Laine} instead of the relativistic ideal gas equation of state. 

\begin{figure}[t]
 \centering
  \includegraphics[width=16cm]{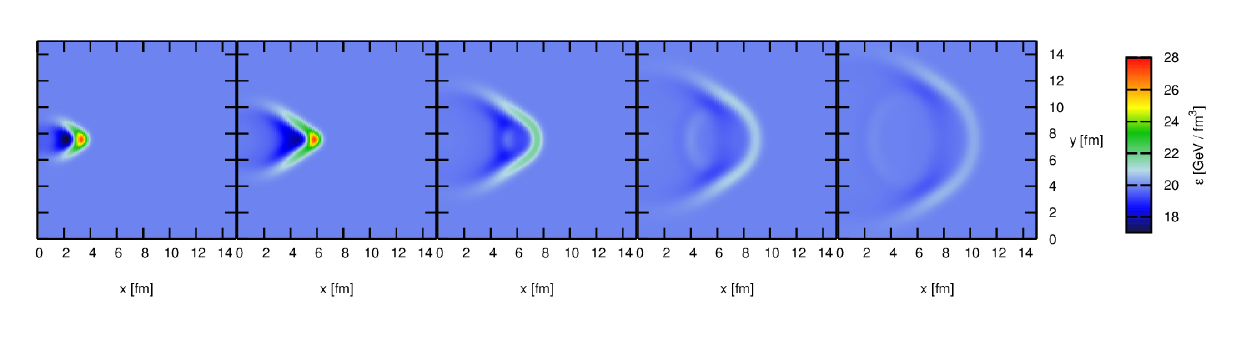}
  \caption{Sequence of energy density profiles during hydrodynamic evolution. 
One hard parton deposits energy and momentum into static medium. It enters from the left. First profile is taken after time delay $t = 2.5$~fm/$c$. 
Each of the following profiles is taken with a delay $\Delta t = 2.5$ fm/$c$ after the 
previous profile. The energy of the parton is fully exhausted after 5 fm/$c$.
The initial energy loss is $dE/dx = -4.148$~GeV/fm, 
initial velocity of the parton is $v_{jet}=0.9999$, and 
unperturbed static energy density is $\epsilon_{0} = 20$ GeV/fm$^{3}$. }
\label{fig:onejet} 
\end{figure}

In the one jet scenario, hard parton deposits energy and momentum into static medium. 
Profiles of the energy density are shown in Fig.~\ref{fig:onejet}; the hard parton enters from left.
The figure displays a sequence of energy density profiles during hydrodynamical evolution. 
We observe a spot with higher energy density at the position where jet deposits energy. 
The increase of the energy density spreads in a Mach-cone-like structure. Behind the jet 
there remains a wake with a dip in the energy density profile. The energy spreads even 
after the jet is fully stopped.  
Equivalent simulation using the ultrarelativistic equation of state ($\epsilon=3p$) leads 
to no significantly different results. Typical energy densities in the simulation are 
far above the transition from hadrons to quarks, where the sensitivity to different equations 
of state is strongest. 
Finally, the evolution of the jet is in qualitative agreement with results presented in \cite{Betz:2008ka}. Especially, 
the explosive burst of energy and momentum deposited by the jet immediately before it 
is fully quenched does not stop the strong
flow behind the jet (the diffusion wake). 

We check that the deposited momentum is almost fully contained in the wake. 
This is done by dividing the space into three regions: (I) is the tube with diameter 
1.5~fm around the jet trajectory up to the shock front of the Mach wave, (II) is the 
rest of the matter behind the cone, and (III) is the region ahead of the shock front 
which is unaware of any energy deposition as yet. At the moment of jet extinction 
region (I) contains 92\% of all jet momentum. After next 2.5~fm/$c$ this drops just a 
little to 87\%. The rest of the momentum is diffused into region (II). This demonstrates
our claim that deposited momentum shows up in form of streams of the hot medium.

\begin{figure}[t]
\centering
\includegraphics[width=16cm]{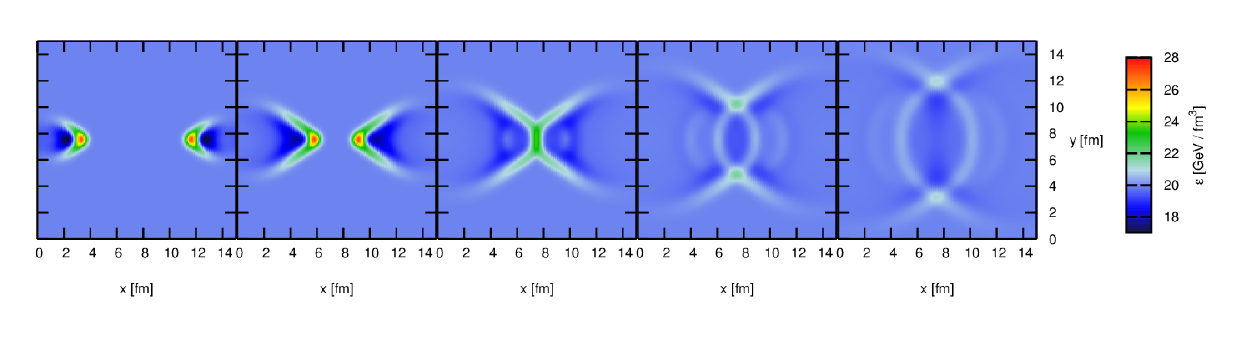}
\caption{Sequence of energy density profiles during hydrodynamical evolution. 
Two hard partons deposit energy and momentum into static medium. They enter from the left
and from the right and move against each other. 
The first profile is taken after time delay $t = 2.5$ fm/$c$. 
Following profiles are taken with subsequent delays of 
$\Delta t = 2.5$~fm/$c$. 
Partons are fully stopped after 5 fm/$c$. Initial energy loss is 
$dE/dx = -4.148$ GeV/fm, initial velocities are $v_{jet}=0.9999$, and 
unperturbed static energy density is $\epsilon_{0} = 20$ GeV/fm$^{3}$.}
\label{fig:jets180} 
\end{figure}
Next, we show results for various scenarios including two hard partons. 
Figure~\ref{fig:jets180} shows hydrodynamic evolution
stimulated by two hard partons moving in opposite directions against each other. 
Both loose the same amount of energy. All energy is deposited into plasma  
before the two partons would meet at $t= 5$ fm/$c$.  
At this time their distance is 3 fm.
Hence,  only the diffusion wakes meet and the two streams of plasma hit each other.  
The cone structures from both hard partons evolve like in previous case with just one 
hard parton. The two streams generated in the wakes meet and stop.

\begin{figure}[t]
\centering
\includegraphics[width=16cm]{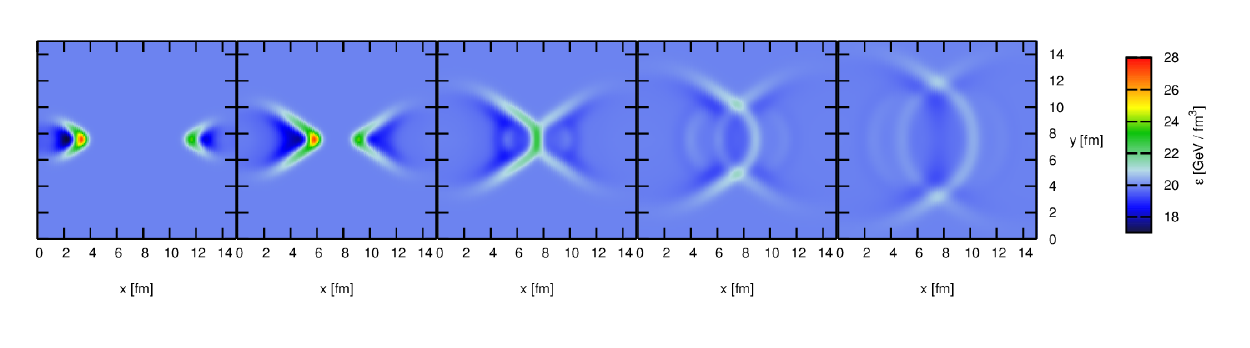}
\caption{Sequence of energy density profiles during hydrodynamic evolution. 
Two hard partons deposit energy and momentum into static medium. 
They enter from the left and from the right and move against each other. 
First profile is taken after time delay $t = 2.5$ fm/$c$ and following profiles
after subsequent delays $\Delta t = 2.5$ fm/$c$. 
Jets are fully quenched after 5 fm/$c$. 
Initial energy loss of the left jet is $dE/dx = -4.148$ GeV/fm. The jet on the right looses one 
half of the energy of the other jet. 
The initial velocity of the jets is $v_{jet}=0.9999$. 
Unperturbed static energy density is $\epsilon_{0} = 20$ GeV/fm$^{3}$. }
  \label{fig:jets180pul} 
\end{figure}
We also examine a setup with a pair of jets in the same directions as before but one of 
them deposits just one half of the energy of the other (Fig. \ref{fig:jets180pul}). 
In total, the jet coming from the left side deposits 21.5~GeV of energy and the opposite 
one 10.5~GeV. 
Again, the Mach cones  pass through each other and continue in their evolution.
\begin{figure}[t]
\centering
\includegraphics[width=9cm]{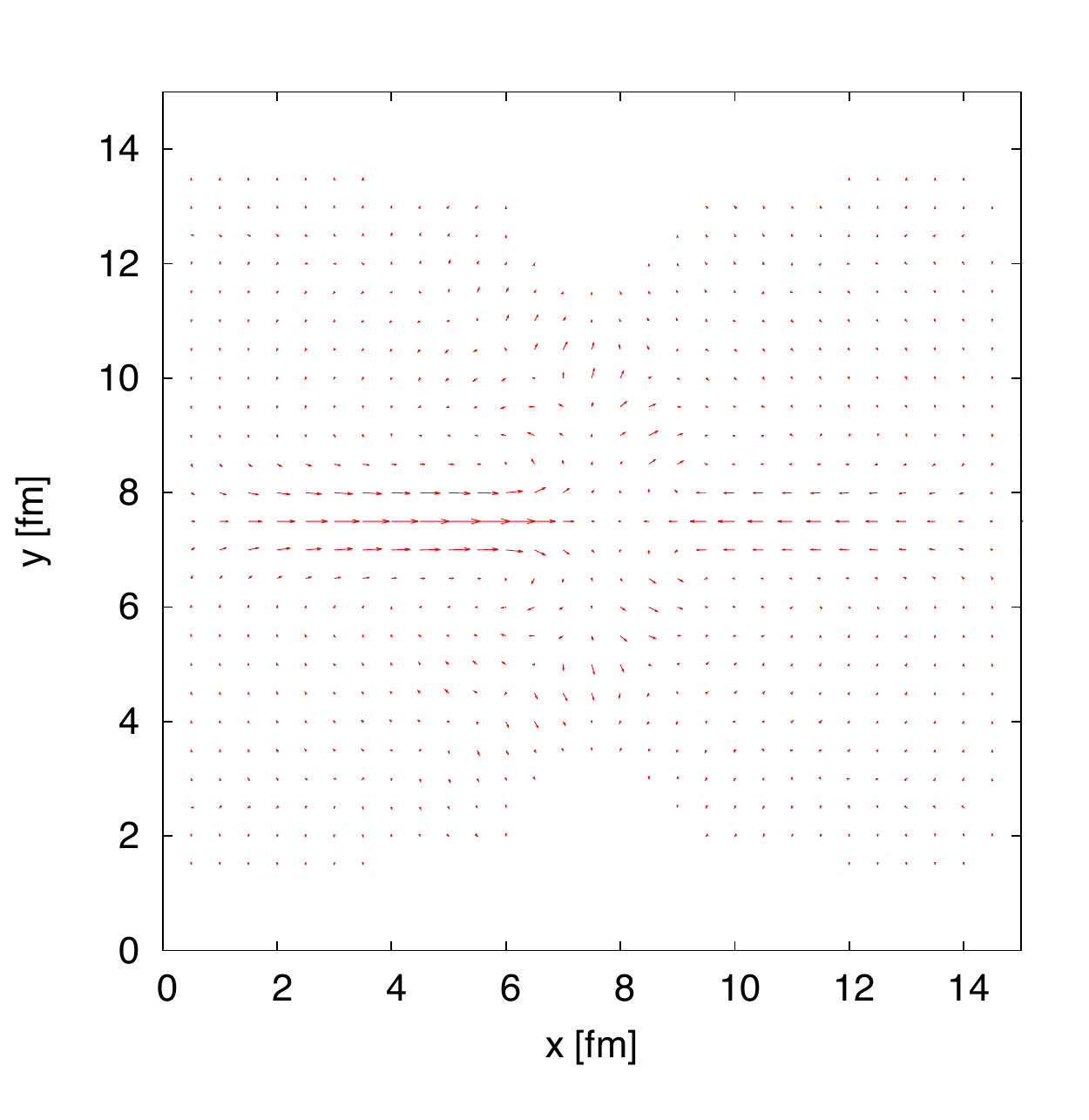}
\caption{Momentum density profile resulting from the situation in Fig.~\ref{fig:jets180pul}.
This profile is taken 10~fm/$c$ after the start of the evolution, i.e.\ 5~fm/$c$ after 
the jets are extinct. The length of the arrows is proportional to the local momentum 
density and their direction shows the direction of the momentum.}
\label{fig:mom180} 
\end{figure}
In Fig.~\ref{fig:mom180} we see that the streams in the wakes continue after the 
jets are extinct and meet slightly closer to the excinction place of the less energetic jet
due to different velocities.
The more energetic streaming, however, does not overturn the streaming on the other side.
\begin{figure}[t]
 \centering
  \includegraphics[width=14cm]{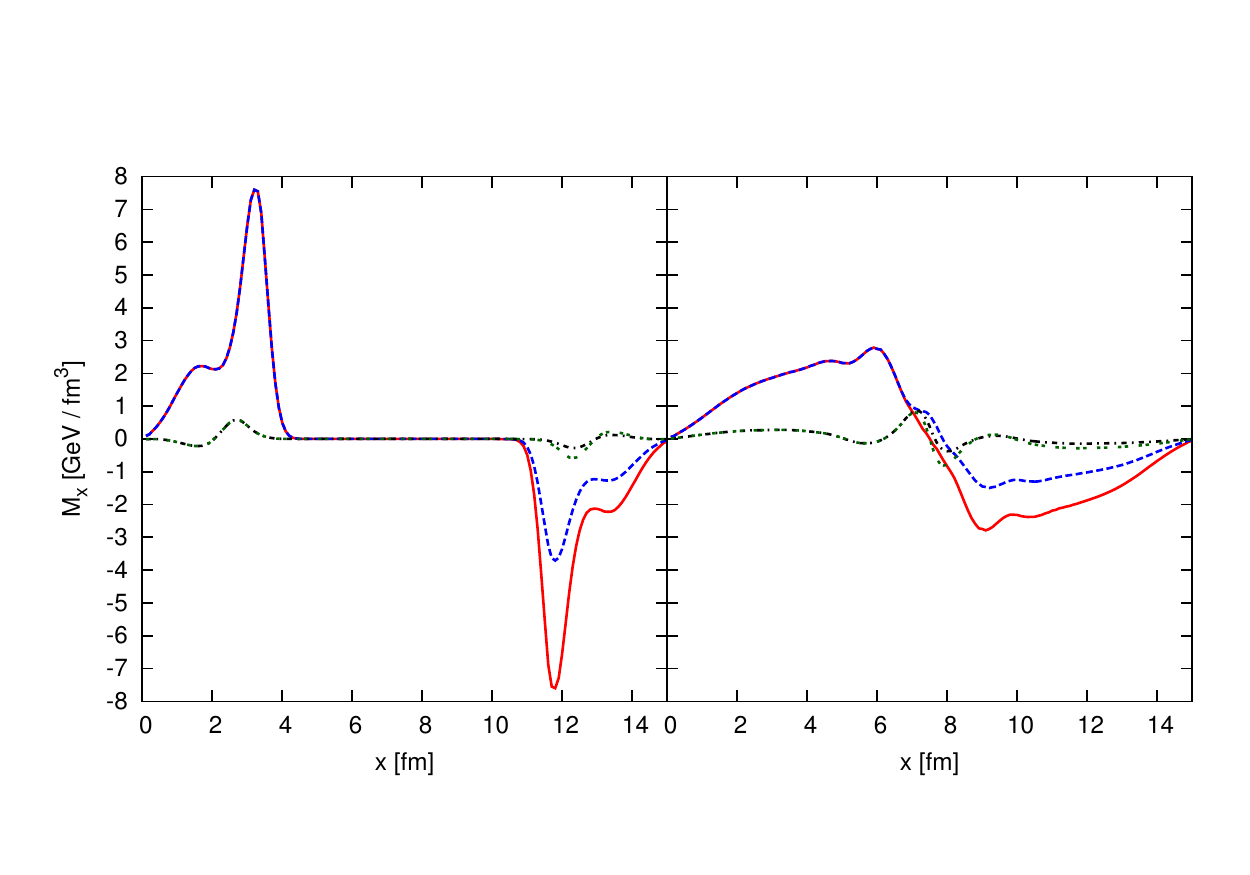}
  \caption{Momentum density along the trajectory of the jets in opposite directions. 
Left panel: momentum density at the moment of extinction. 
Right panel: momentum density after another 2.5 fm/$c$. 
Red solid line: pair of jets with equal energy.
Blue dashed line: pair of jets where jet on the right has originally a half of the other 
jet's energy. 
Green double dashed line: pair of jets in opposite directions, jet on the right is moving along the x-axis in the axial distance of 2 fm from the other jet's trajectory; momentum density distribution is taken along the line in centre between the two jets. 
Black dash and dotted line: same as previous case but jet entering from right has originally a half of the other jet's energy.}
\label{fig:jets180mom}
\end{figure}
Deposited momentum for various scenarios with jets aiming in opposite directions 
is also shown in Fig.~\ref{fig:jets180mom}. This figure confirms that the momentum 
density does not overrun to the other side even if one jet deposits more momentum than 
the other (blue dashed line). If the two jets do not collide exactly head-on, momentum density 
along the line between their trajectories shows the vortex structures \cite{Betz:2008ka}: 
close to the position of the  parton there is
momentum in the direction of the parton and shortly after it momentum turns into the other 
direction. This is the vortex where matter flows backwards from the Mach cone to the wake
where the energy density and pressure is lower than in the unperturbed medium.

\begin{figure}[t]
\centering
\includegraphics[width=16cm]{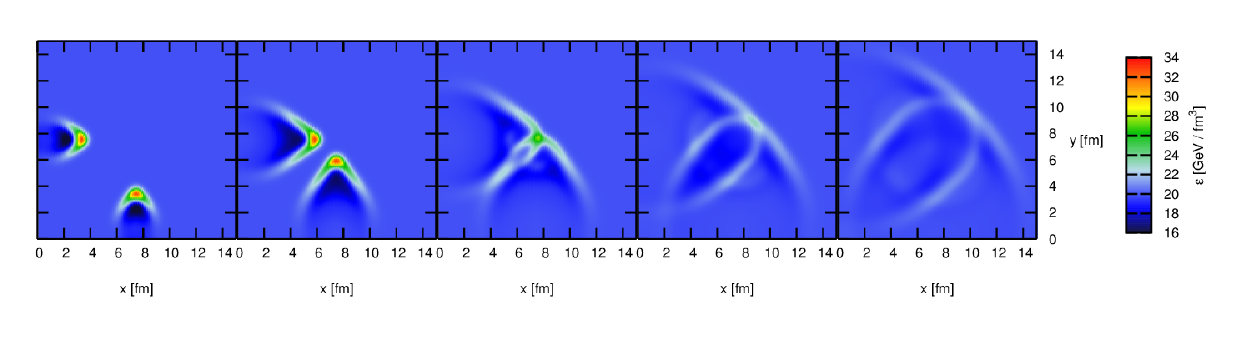}
\caption{Sequence of energy density profiles during  hydrodynamical evolution. 
Two hard partons deposit energy and momentum into static medium. One enters from the left, one enters from the bottom. 
First profile is taken after a delay of  $t = 2.5$~fm/$c$, the subsequent profiles after time steps
$\Delta t = 2.5$ fm/$c$. Jets are fully quenched after 5 fm/$c$. 
The initial energy loss is $dE/dx = -8.296$~GeV/fm, initial velocity is $v_{jet}=0.9999$,
unperturbed static energy density is $\epsilon_{0} =20$ GeV/fm$^{3}$. }
  \label{fig:jets90} 
\end{figure}
\begin{figure}[t]
 \centering
 \includegraphics[width=16cm]{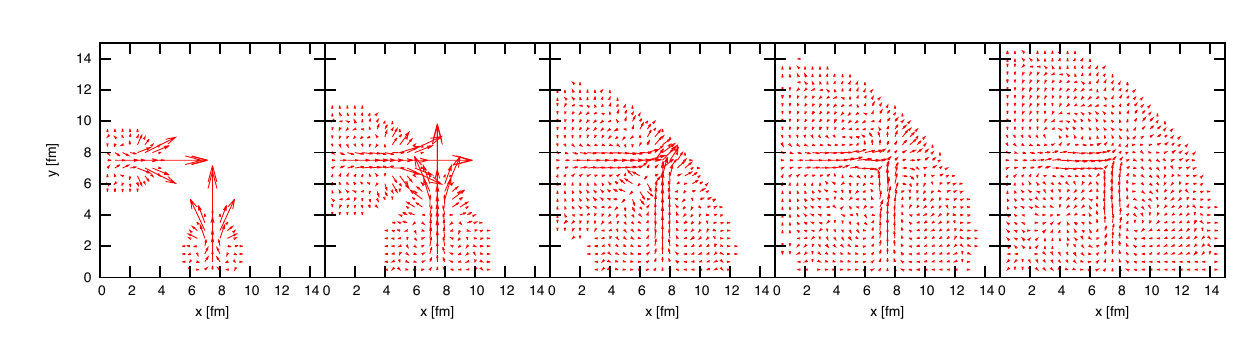}
 \caption{Situation from Fig.~\ref{fig:jets90}. The direction and length of arrows 
correspond to the direction and magnitude of local momentum density, respectively.}
  \label{fig:momentaprofile}
\end{figure}
In  real situation, jet-induced streams will come together under various angles. In order 
to see how they may interact we examine a situation where the two jets move perpendicularly.
In Fig.~\ref{fig:jets90} two hard partons---one entering from the left and one  from 
the bottom in the same plane---deposit energy and momentum into static medium. The 
unperturbed momentum density here is again $\epsilon_{0}=20$ GeV/fm$^{3}$. In order to see
a stronger effect on the medium, energy-momentum deposition was doubled in comparison 
with previous examples, e.g. $dE/dx = -8.296$ GeV/fm. 
Jets are fully stopped  1.5 fm/$c$ before they would meet. We see that the conical structures
of higher energy density look like a superposition of the two Mach cones that propagate 
also after the full quenching of the jets. In order to study the induced streaming of matter we
plot the evolution of mometum density profile in Fig.~\ref{fig:momentaprofile}. We can 
observe the merging of the two wake streams in the place where they come togehter. 
If energy-momentum deposition of the jet into medium is sufficient, 
on a small distance---until the energy is diffused too much---the streams continue 
together diagonally.
Vortex-like structure in momentum density vectors in medium is present \cite{Betz:2008ka}. 

\begin{figure}[h]
 \centering
  \includegraphics[width=14cm]{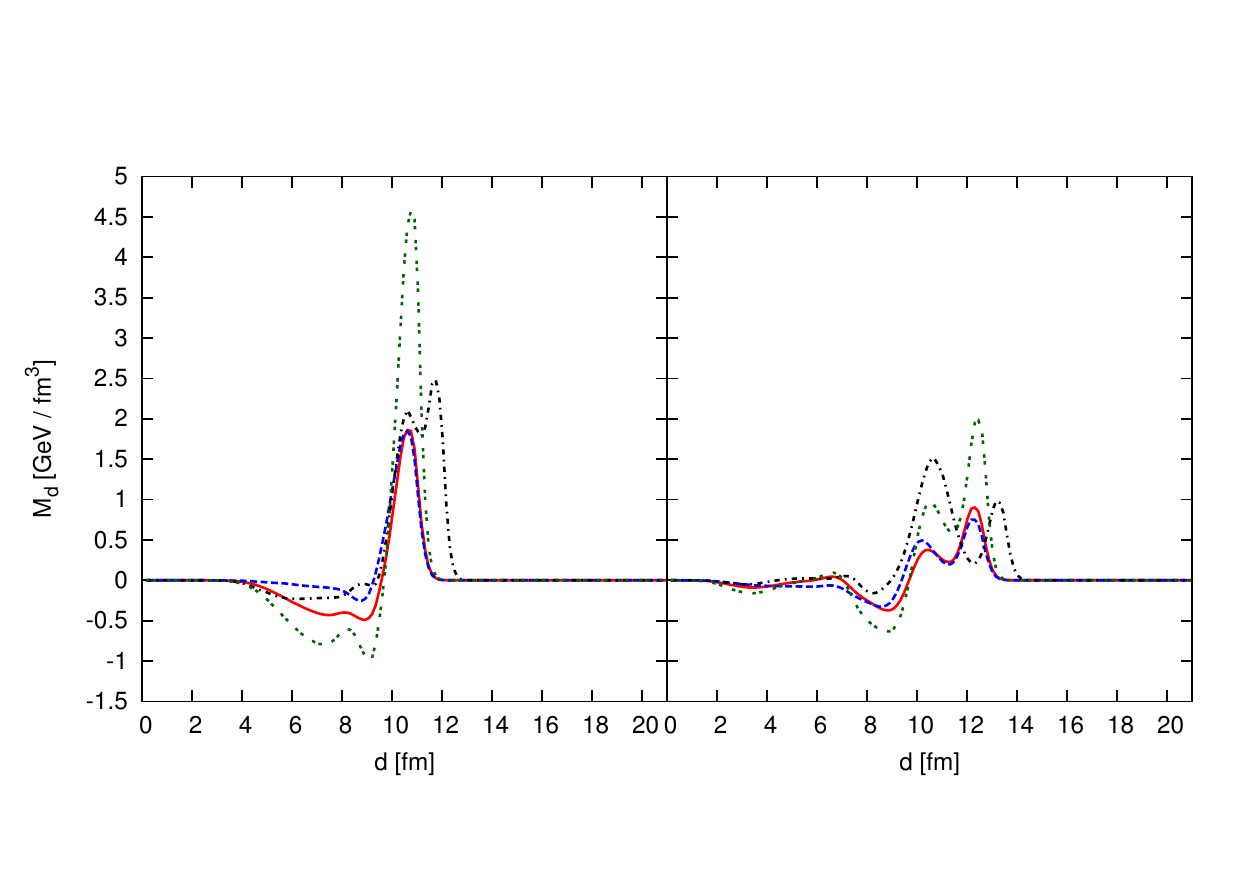}
  \caption{Momentum density in diagonal direction during hydrodynamic evolution
as a function of the diagonal coordinate. Jets move perpendicularly to each other and
source terms vanish at $t = 5$~fm/$c$, which is 1.5 fm/$c$ 
before they would meet, unless stated othewise.
Left panel: profile at $t = 7.5$ fm/$c$, right panel:   $t = 10.0$ fm/$c$. 
Red solid line: perpendicular jets scenario, both jets have equal energy. 
Blue dashed line: same scenario as before but the lower side jet has  
a half of the other jet's energy. 
Green double dashed line:  each jet deposits two times more energy than in the first case. 
Black dash and dotted line:  jets source terms vanish only 0.5 fm/c before jets would meet.}
\label{fig:jets90mom1}
\end{figure}
The merging of diffusion wakes is also demonstrated in Fig. \ref{fig:jets90mom1}. 
The figures display  momentum density in the direction diagonal  to the two jets as a function 
of the diagonal coordinate for various perpendicular jets scenarios. 
Left panel displays the situation when the wakes just merge ($t=7.5$~fm/$c$). 
Right panel shows the situation after another 2.5~fm/$c$ ($t=10.0$ fm/$c$). 
Although the jet source terms vanish  before the wakes make contact, the merged wakes 
continue to evolve. Momentum density also exhibits double peak structure. 
Behind the merged wakes the momentum density turns negative, i.e.~it points 
backwards. This is a part of the vortices that are built up on the sides around the jets. 
We observe that the lower unperturbed energy density or higher energy-momentum 
deposition induce higher momentum density
on the diagonal when wakes merge, as expected. 
The peaks in momentum density corresponding to merging 
of two wakes with equal and also with unequal 
energy seem qualitatively similar.

\begin{figure}[t]
\centering
\includegraphics[width=16cm]{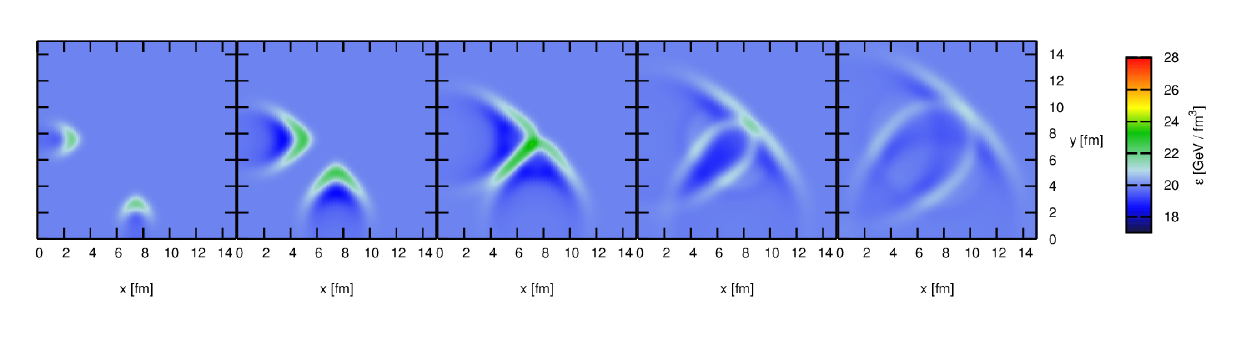}
\caption{Sequence of energy density profiles during a hydrodynamical evolution. 
Two hard partons deposit energy and momentum into static medium. One enters from the left, one enters from the 
bottom. Distance of closest approach of extrapolated trajectories is 2 fm. Profiles are taken 
on the plane in the middle of both trajectories. First profile is taken after $t = 2.5$ fm/$c$. 
All other profiles are taken with subsequent delays $\Delta t = 2.5$ fm/$c$. 
Jets are quenched after 5 fm/$c$, i.e.\ in the second figure.
The initial energy loss is $dE/dx = -4.148$ GeV/fm, initial velocity is $v_{\mathrm jet}=0.9999$, 
unperturbed static energy density is $\epsilon_{0}=20$ GeV/fm$^{3}$.}
\label{fig:jets90b=2fm}
\end{figure}

It is very rare, however, that two jets would be aimed so precisely that their wakes
meet  exactly as it was assumed here. Therefore, we examine a situation with velocities 
perpendicular to each other but the distance of closest approach of their extrapolated 
trajectories is 2~fm. The evolution of the energy density on the plane in the middle
between the two trajectories is shown in Fig.~\ref{fig:jets90b=2fm}.  
We checked also the plots of velocity and momentum densities. The wake streams
and their merging is less visible  in such a distance from the original jets. On the other hand, 
one better sees the vortices that are built at the two sides behind the jet. 

Diagonal dependences of the momentum density look in this case qualitatively 
similarly as in Fig.~\ref{fig:jets90mom1}, only the size of momentum densities is
typically lower.


\section{Summary and conclusions}
\label{s:summ}

We presented simulations of energy and momentum deposition from hard 
partons into hot deconfined matter and studied the response of the matter. 
The aim of this toy model simulation was to verify  
that the streams which are generated in the wakes indeed carry the deposited 
momentum. It remains to be checked to what extent this feature may be spoiled 
by introducing momentum transport parametrised by viscosity into the simulation
\cite{Bouras:2012mh}. 
Moreover, when two streams meet and are not being stimulated anymore by 
a hard parton, they merge into one stream which continues until it is tamed by diffusion. 
We thus confirmed the assumption made in \cite{Tomasik:2011xn}
that the generated wakes interact and influence each other. That paper 
concluded that under such conditions in  heavy ion collisions 
isotropically produced hard partons 
generate interacting wakes which can lead to collective motion that exhibits elliptic flow
correlated with the direction of the reaction plane. 
It was estimated to be on the level of one or two per cent there. 

Hence, we see that we have identified a mechanism which generates anisotropic 
flow and consequently anisotropic particle production in ultrarelativistic nuclear 
collisions. Let us stress again the difference to the hydrodynamic scenario with hot spots in 
the initial conditions. That model takes into account that the initial energy density profile
for hydrodynamic simulation changes event-by-event. In contrast to our model all 
features unique for a given event are specified in the initial conditions from which 
hydrodynamics based on energy-momentum conservation is started. In our case, 
energy and momentum is deposited into the medium \emph{during} its evolution
and not only at the beginning. 

The influence of hard partons on flow anisotropies will show itself in various orders 
of particle production anisotropies. It remains to be studied, how big the effect
is in realistic simulations reproducing the dynamics of the fireball. Also, inclusion 
of viscosity will weaken the influence of this effect, but quantitative details remain to 
be investigated. When analyzing 
precision data from LHC (and RHIC) it will be important, however, 
to take this effect into account 
if one wants to make quantitative statements about the transport properties of the hot
deconfined matter.


\paragraph{Acknowledgements}
This work was partially supported by RVO68407700 (Czech Republic).
BT also acknowledges support by VEGA~1/0457/12 and APVV-0050-11 (Slovakia).


\end{document}